\documentclass[epj]{svjour}
\usepackage[dvips]{graphicx}
\begin{document}
\title{Consequences of temperature fluctuations in
observables measured in high energy collisions}
\author{Grzegorz Wilk \inst{1}
\thanks{\emph{e-mail: wilk@fuw.edu.pl}}
\and Zbigniew W\l odarczyk\inst{2}
\thanks{\emph{e-mail: zbigniew.wlodarczyk@ujk.kielce.pl}}
}                     
\institute{National Centre for Nuclear Research,
        Department of Fundamental Research, Ho\.za 69, 00-681
        Warsaw, Poland
         \and Institute of Physics,
                Jan Kochanowski University,  \'Swi\c{e}tokrzyska 15,
                25-406 Kielce, Poland}
\date{Received: date / Revised version: date}
%
\abstract{ We review the consequences of intrinsic, nonstatistical
temperature fluctuations as seen in observables measured in high
energy collisions. We do this from the point of view of
nonextensive statistics  and Tsallis distributions. Particular
attention is paid to multiplicity fluctuations as a first
consequence of temperature fluctuations, to the equivalence of
temperature and volume fluctuations, to the generalized
thermodynamic fluctuations relations allowing us to compare
fluctuations observed in different parts of phase space, and to
the problem of the relation between Tsallis entropy and Tsallis
distributions. We also discuss the possible influence of
conservation laws on these distributions and provide some examples
of how one can get them {\it without} considering temperature
fluctuations.
} 
\authorrunning{G. Wilk and Z. W\l odarczyk}
\titlerunning{Consequences of temperature fluctuations ...}

\maketitle
\section{Introduction}
\label{sec:Introduction}

Nowadays the statistical approach is a standard procedure used to
model high energy multiparticle production processes
\cite{MG_rev}. However, it has been realized that data on many
single particle distributions deviate in a visibly way from what
one expects from the usual statistical models, based on
Boltzman-Gibbs (BG) statistics. These frequently show power-like
rather than exponential behavior, and, in addition, multiparticle
distributions are broader than naively expected. These
observations prompted the idea of a suitable modification of a
simple statistical approach used by including in it the
possibility of accounting for possible intrinsic, nonstatistical
fluctuations. These were identified as the source of the
deviations. Such fluctuations are important as possible signals of
phase transition(s) taking place in an hadronizing system
\cite{PhTr}. Therefore it is important to be able to include them.
In this way the Tsallis statistical approach \cite{Tsallis},
already known in other branches of physics, was successfully
introduced to the field of multiparticle production
processes\footnote{For details see our previous review
\cite{WWrev}. Here we present recent developments in this field
not covered there.}. In this approach a new parameter, the
nonextensivity parameter $q$ appears, which is identified with
fluctuations of the parameter $T$ identified with the
"temperature" of the hadronizing fireball \cite{WWq}.

It was shown there that such a situation can only occur when the
heat bath is not homogeneous and must be described by a local
temperature, $T$, fluctuating from point to point around some
equilibrium value, $T_0$. Assuming some simple diffusion picture
as being responsible for equalization of this temperature
\cite{WWq,WWrev} one obtains the evolution of $T$ in the form of a
Langevin stochastic equation with the distribution of $1/T$,
$g(1/T)$, emerging as a solution of the corresponding
Fokker-Planck equation. It turns out that in this case $g(1/T)$
takes the form of a gamma distribution,
\begin{eqnarray}
g(1/T) &=& \frac{1}{\Gamma\left(\frac{1}{q -
1}\right)}\frac{T_0}{q - 1}\left(\frac{1}{q -
1}\frac{T_0}{T}\right)^{\frac{2 - q}{q - 1}}\cdot\nonumber\\
&& \cdot \exp\left( - \frac{1}{q - 1}\frac{T_0}{T}\right).
\label{eq:gamma}
\end{eqnarray}
Convoluting the usual Boltzman-Gibbes exponential factor $\exp (-
E/T)$ with this $g(1/T)$, one immediately gets a Tsallis
distribution, $h_q(E)$, with a new parameter $q$, which for $q
\rightarrow 1$, becomes the usual BG distribution\footnote{Notice
that all distributions used here are defined as probability
density functions with standard normalization, $\int dE h_q(E) =
1$. This results in the presence of the prefactor $(2-q)/T$.}:
\begin{eqnarray}
h_q(E) &=& \frac{2-q}{T}\exp_q \left(-\frac{E}{T}\right)
=\nonumber\\
&=& \frac{2-q}{T}\left[1 - (1-q)\frac{E}{T}\right]^{\frac{1}{1-q}}
 \label{eq:Tsallis}\\
 && \stackrel{q \rightarrow 1}{\Longrightarrow}\, \frac{1}{T}\exp
\left(-\frac{E}{T}\right). \label{eq:BG}
\end{eqnarray}
with
\begin{equation}
q = 1 + \omega_T^2\quad {\rm where}\quad \omega^2_T =
\frac{Var(T)}{<T>^2}, \label{eq:q}
\end{equation}
directly connected to the variance of $T$. This idea was further
developed in \cite{BJ} and \cite{B} (where problems connected with
the notion of temperature in such cases were addressed). This
forms a basis for so-called {\it superstatistics} \cite{SuperS}.
In what follows, we shall use this approach when discussing
Tsallis distributions (except of Section \ref{sec:Tsallis} in
which we compare it with distribution obtained from Tsallis
entropy).

\begin{figure}[h]
\begin{center}
\resizebox{0.45\textwidth}{!}{
  \includegraphics{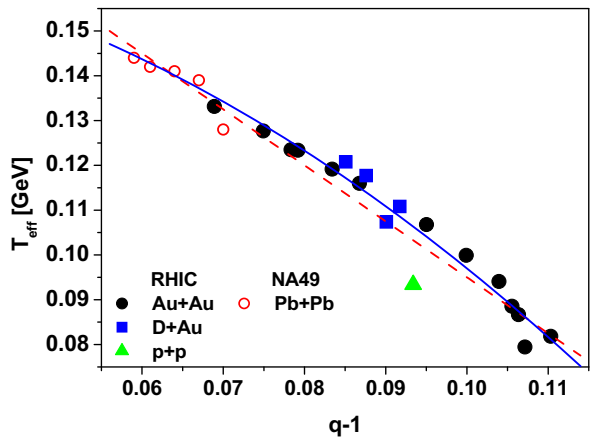}
  }
\caption{(Color online) Dependence of $T_{eff}$ on $q$ for
different energies. RHIC data points are from \cite{BD} whereas
NA49 points are from \cite{NA49} (for, respectively, $\sqrt{s} =
6.3,~7.6,~8.8,~12.3,~17.3$ GeV (negative pions). Fits are:
$T_{eff} = 0.17 - 7.3(q - 1)^2$ (full line, and $T_{eff} = 0.22 -
1.25(q - 1)$ (dashed line). In both cases $T_{eff}$ is in GeV.}
\label{Fig_T_v_q}
\end{center}
\end{figure}

It must be mentioned that temperature fluctuations (visualized by
$q > 1$ values of the nonextensivity parameter) also allow for a
description of the possible energy transfer from or to the heat
bath \cite{WWrev}. Namely, if $T_v$ is a new parameter
characterizing such an energy transfer, then
\begin{equation}
T \rightarrow T_{eff} = T_0 + (q - 1) T_v, \label{eq;T_eff}
\end{equation}
Fig. \ref{Fig_T_v_q} shows that such an effect is indeed observed
\cite{WWprc}. It is caused mainly by the possible energy transfer
between the central fireball (participants) and nuclear fragments
passing by without interaction (spectators)\footnote{Similar
effect is also expected in propagation of cosmic rays through the
outer space, cf., \cite{qCR}; we shall not discuss this issue
here.}. Notice that this energy transfer is only possible in the
presence of fluctuations, i.e., for $ q
> 1$, when there are no fluctuations and $q = 1$ one has $T_{eff} = T_0$.

It is worth mentioning at this point that fluctuation phenomena as
discussed here can be incorporated into a traditional presentation
of thermodynamics \cite{M}. In such a general approach, the
Tsallis distribution (\ref{eq:Tsallis}) belongs to the class of
general admissible distributions which satisfy thermodynamic
consistency conditions and present a natural extension of the
usual BG canonical distribution (\ref{eq:BG}). This, together with
a recent generalization of classical thermodynamics to a
nonextensive case presented in \cite{recentB}, form a constructive
answer to the critical remarks we encountered concerning the
consistency of Tsallis statistics with the usual thermodynamics in
\cite{debate}.

Applications of Tsallis distributions to multiparticle production
processes are now numerous. To those quoted previously in
\cite{WWrev} one should add some new results from
\cite{WWprc,BPU_epja,WWrev1} and presented in \cite{B}. The most
recent applications of this approach come from the STAR and PHENIX
Collaborations at RHIC \cite{STAR,PHENIX} and from CMS \cite{CMS},
ALICE \cite{ALICE} and ATLAS \cite{ATLAS} Collaborations at LHC
(see also a recent compilation \cite{qcompilation})\footnote{In
addition to applications presented in this review, the
nonextensive approach has also been applied to hydrodynamical
models \cite{OsadaW} and to investigations of dense nuclear matter
\cite{RW}.}. In Section \ref{sec:Imprints} we report on new
results concerning the consequences of temperature fluctuations in
observables measured in high energy collisions obtained since our
previous review \cite{WWrev}. In Section \ref{section:Conditional}
the influence of conservation laws, forcing the use of conditional
probabilities and resulting in $q < 1$, is discussed. In Section
\ref{sec:Tsallis}, the differences between Tsallis distributions
as obtained from Tsallis entropy and the concept of
superstatistics is discussed. A possible experimental check is
proposed. Section \ref{section:nonthermal} is devoted to yet
another, not based on statistical models, derivation of Tsallis
distribution. Section \ref{sec:Summary} is our summary.

\section{Imprints of superstatistic in multiparticle
processes} \label{sec:Imprints}

\subsection{Multiplicity distributions}
\label{sec:Multiplicity}

In \cite{WWcov} (cf. also \cite{WWrev}) we saw that $T$
fluctuations in the form of Eq. (\ref{eq:gamma}) not only result
in power-like behavior of single particle distributions, but also
in a specific broadening of the corresponding multiplicity
distributions, $P(N)$, which evolve from the poissonian form
characteristic of BG distributions to the negative binomial (NB)
form for Tsallis distributions. In short:  whenever we have $N$
independently produced secondaries with energies $\{
E_{i=1,\dots,N}\}$ taken from the exponential distribution $f(E)$,
cf. Eq. (\ref{eq:BG}), in which case the corresponding joint
distribution is given by
\begin{equation}
f\left( \{ E_{i=1,\dots,N}\}\right) =
\frac{1}{\lambda^N}\exp\left( - \frac{1}{\lambda}\sum^{N}_{i=1}
E_i\right), \label{eq:PNBG}
\end{equation}
and whenever
\begin{equation}
\sum^N_{i=0} E_i \leq E \leq \sum^{N+1}_{i=0} E_i,
\label{eq:condition}
\end{equation}
then the corresponding multiplicity distribution is poissonian,
\begin{equation}
P(N) = \frac{\left( \bar{N}\right)^N}{N!} \exp\left ( -
\bar{N}\right) \quad {\rm where}\quad \bar{N} =\frac{E}{\lambda}.
\label{eq:Poisson}
\end{equation}
But whenever in a given process $N$ particles with energies $\{
E_{i=1,\dots,N}\}$ are distributed according to the joint
$N$-particle Tsallis distribution,
\begin{equation}
h\left(\{ E_{i=1,\dots,N}\} \right) = C_N\left[ 1-
(1-q)\frac{\sum^N_{i=1} E_i }{\lambda} \right]^{\frac{1}{1-q}+1-N}
\label{eq:NTsallis}
\end{equation}
(for which the corresponding one particle Tsallis distribution
function in Eq. (\ref{eq:Tsallis}) is the marginal distribution),
then, under the same condition (\ref{eq:condition}), the
corresponding multiplicity distribution is the NB distribution
\cite{PN},
\begin{eqnarray}
P(N) &=& \frac{\Gamma(N+k)}{\Gamma(N+1)\Gamma(k)}\frac{\left(
\frac{\langle N\rangle}{k}\right)^N}{\left( 1 + \frac{\langle
N\rangle}{k}\right)^{(N+k)}},\label{eq:NBD}\\
&& {\rm where}\quad k=\frac{1}{q-1}.\nonumber
\end{eqnarray}
For $q\rightarrow 1$ one has $k\rightarrow \infty$ and
(\ref{eq:NBD}) becomes a poissonian distribution
(\ref{eq:Poisson}), whereas for $q\rightarrow 2$ one has
$k\rightarrow 1$ and (\ref{eq:NBD}) becomes a geometrical
distribution. For large values of $N$ and $\langle N\rangle$ Eq.
(\ref{eq:NBD}) can be written in the following scaling form,
\begin{equation}
\langle N\rangle P(N) \cong  \psi\left( z=\frac{N}{\langle
N\rangle} \right) = \frac{k^k}{\Gamma(k)} z^{k-1}\exp( - kz),
\label{eq:scalingform}
\end{equation}
known as Koba-Nielsen-Olesen (KNO) scaling
\cite{KNO,GOR}\footnote{The connection between $q$ and $k$ was
first discovered when fitting $p\bar{p}$ data for different
energies by means of the Tsallis formula (\ref{eq:Tsallis})
\cite{RWW}). The resulting energy dependence of parameter $q$
turned out to coincide with that of $1/k$ from the respective NB
distribution fits to the corresponding $P(N)$. It was then
realized that fluctuations of $\bar{N}$ in the poissonian
distribution (\ref{eq:Poisson}) taken in the form of
$\psi(\bar{N}/<N>)$, Eq. (\ref{eq:scalingform}), lead to the NB
distribution (\ref{eq:NBD}).}.

Note that, if in the Poisson distribution (\ref{eq:Poisson}) one
fluctuates the mean value, $\bar{N} = E/T$ (valid for
one-dimensional, $D = 1$, case), using its distribution in the
form
\begin{equation}
g\left( \bar{N}\right) = g\left( \frac{1}{T}= \frac{\bar{N}}{E}
\right)\left| \frac{d\bar{N}}{d(1/T)}\right|, \label{eq:gbarN}
\end{equation}
(where $g(1/T)$ is given by Eq. (\ref{eq:gamma})) then the
resulting multiplicity distribution
\begin{equation}
P(N) = \int d\bar{N} g\left( \bar{N}\right) \frac{\bar{N}}{N!}
\exp\left( - \bar{N}\right) \label{eq:poisgam}
\end{equation}
is the NB distribution given by Eq.
({\ref{eq:NBD})\footnote{Actually this has been also noted in
\cite{Shih,B} and recently discussed in \cite{VP} where the credit
in what concerns the origin of discussion of such connection
between the Poisson and NB distributions has been given to
\cite{RAF}.}.

\subsection{Equivalence of temperature fluctuations and
volume fluctuations}
\label{sec:Temperature}

The KNO scaling form (\ref{eq:scalingform}), with assumed
identification
\begin{equation}
z = \left( \frac{V}{\langle V\rangle} \right)^{1/4}, \label{eq:TV}
\end{equation}
(where $V$ is the volume of the interaction region) has been used
in \cite{Vfluct} as a starting point for a description of particle
spectra by means of {\it fluctuations of volume}. In this way it
was hoped to avoid the notion of {\it fluctuating temperature}
discussed here. The results were encouraging. However, for
constant total energy as assumed in \cite{Vfluct}, $E=const$, both
the volume $V$ and temperature $T$ are related,
\begin{equation}
E \sim VT^4, \label{eq:VT}
\end{equation}
this means that
\begin{equation}
T = \langle T\rangle\left( \frac{\langle V \rangle}{V}
\right)^{\frac{1}{4}} \label{eq:TV}
\end{equation}
and the mean multiplicity in the microcanonical  ensemble (MCE),
$\bar{N}$, can be written as
\begin{equation}
\bar{N} = \langle N\rangle\cdot \frac{V}{\langle V\rangle}\left(
\frac{T}{\langle T\rangle}\right)^3 = \langle N\rangle
\frac{\langle T\rangle}{T}. \label{eq:barNV}
\end{equation}
This implies that both approaches are equivalent and that
fluctuations of $V$ {\it assumed} in \cite{Vfluct} in the form
given by Eq. (\ref{eq:TV}) arise as an effect of fluctuations of
$T$ considered here with $g(1/T)$ given by Eq. (\ref{eq:gamma}).
This is not assumed but {\it derived} from the properties of the
underlying physical process in the nonhomogeneous heat bath. One
should also remember that UA5 data \cite{UA5} show that KNO
scaling is broken due to the energy dependence of the parameter
$k$\footnote{A possible solution to solve the breakdown of the KNO
scaling in multiplicity distributions measured in $e^+e^-$ and
$pp$ collisions has been proposed in \cite{R1}.}. In fact, as
shown in \cite{PN}, $k^{-1} = -0.104 + 0.058\ln \sqrt{s}$ .
Therefore, in the scenario with fluctuations of the volume $V$,
the scaling KNO form of the $P(N)$ used to model these
fluctuations is a somewhat rough simplification. On the contrary,
in the scenario of temperature $T$ fluctuations, $ P(N)$ is given
by a NB distribution, which adequately describes the data.

\subsection{Relation between fluctuations observed in different parts of phase
   space}
\label{sec:Relation}

\subsubsection{$q$-sum rules}
\label{sec:Sumrules}

So far, fluctuations of $T$ as introduced in \cite{WWq} and
measured by the corresponding parameter $q$ were discussed using
examples of distributions of longitudinal phase space (in the
rapidity  variable $y$ and integrated over transverse momenta),
$dN/dy$, and in transverse phase space, $dN/d\vec{p}_T$. It was
found that the corresponding parameters $q$, $q = q_T$ and $q =
q_L$, respectively, are different. Whereas $q_L - 1 \sim 0.1 -
0.3$ and grows with the energy of collision (measured mainly in
$pp$ and ${\bar p}p$ collisions), transverse fluctuations are much
weaker, $q_T - 1 \sim 0.01-0.1$ and vary slowly with energy
(depending  only slightly on whether one observes elementary
collisions or collisions between nuclei) \cite{QLQT,MB}. As shown
in \cite{WWcov,WWrev} the same fluctuations of $T$ result in
broadening of multiplicity distributions resulting in its NB form
as given by Eq.({\ref{eq:NBD}). This time the corresponding $q$
describes fluctuations in the whole of phase space, with $p =
\sqrt{|\vec{p}^2|} = \sqrt{p^2_L + p_T^2}$.

In \cite{QLQT} it was proposed that, because $q-1 =
\sigma^2(T)/\langle T\rangle^2$ (i.e., is given by fluctuations of
the total temperature $T$), and assuming that $\sigma^2(T) =
\sigma^2(T_L) + \sigma^2(T_T)$, the resulting values of $q$ should
not be too different from
\begin{equation}
q\, =\, \frac{q_L\langle T_L\rangle^2 + q_T\langle
T_T\rangle^2}{\langle T\rangle^2} - \frac{\langle T_L\rangle^2
+\langle T_T\rangle^2}{\langle T\rangle^2} + 1 . \label{eq:qqq}
\end{equation}
Therefore, because of the dominance of longitudinal (partition)
temperature over transverse, $T_L \gg T_T$, one should expect that
$q \sim q_L$. This is indeed observed \cite{QLQT}. This is the
first sum rule for parameters $q$ obtained from different
measurements.

Fluctuations of temperature are usually deduced either from data
averaged over all other possible fluctuations or from data also
accounting for fluctuations of other measured variables. In this
case one can refine the experimentally evaluated $q$ and, for
example, when extracting $q$ from distributions of $dN/dy$, one
finds that (cf., \cite{WWW} for details)
\begin{equation}
q - 1 \stackrel{def}{=} \frac{Var(T)}{\langle T\rangle^2} =
\frac{Var(z)}{\langle z\rangle^2} - \frac{Var\left(
m_T\right)}{\langle m_T \rangle^2}, \label{eq:sumrule}
\end{equation}
where $z = m_T/T$ (with $m_T = \sqrt{m^2 + p_T^2}$). This is the
second sum rule for the nonextensivity parameters $q$ obtained
from different measurements. It connects total $q$, which can be
obtained from an analysis of the NB form of the measured
multiplicity distributions, $P(N)$, with $q_L -1 = Var(z)/\langle
z\rangle^2$, obtained from fitting rapidity distributions and
$Var\left(m_T\right)/\langle m_T\rangle^2$ obtained from data on
transverse mass distributions. When extracting $q$ from
distributions of $dN/dm_T$, we proceed analogously but now with
$z=\cosh y/T$.

\subsubsection{Generalized thermodynamic fluctuation relations}
\label{sec:Generalized}

So far, we concentrated only on fluctuations of $T$. We shall
continue the discussion by allowing the energy ($U$), temperature
($T$) and multiplicity ($N$) of the system to fluctuate and
propose to express these fluctuations by the corresponding
parameter $q$ \cite{GTR}. Our discussion is based on the notion of
thermodynamic uncertainty relations discussed in \cite{BH}. It was
suggested there that the temperature $T$ and energy $U$ could be
regarded as complementary, similarly as are energy and time in
quantum mechanics. One expects from simple dimensional analysis
that ($k$ is Boltzman's constant)
\begin{equation} \Delta U\, \Delta
\beta \ge k ,\quad {\rm where}\quad \beta = 1/T \label{eq:DUDB}
\end{equation}
Definite $U$ (isolation) and definite $T$ (contact with a heat
bath) to represent the two extreme cases of this complementarity.
This leads to the so called Lindhard's uncertainty relation
between the fluctuations of $U$ and $T$ \cite{L}\footnote{This
idea is still disputable, see \cite{L}, nevertheless we shall
treat these increments as a measure of fluctuations of the
corresponding physical quantities.}:
\begin{equation}
\omega_U^2\, +\, \omega^2_T\, =\, \frac{1}{\langle
N\rangle}\qquad{\rm where} \qquad  \omega^2_x = Var(x)/\langle
x\rangle^2 , \label{eq:JL}
\end{equation}
and this, as was shown in \cite{GTR}, can be generalized to
include all variables: $U$, $T$ and $N$ by using the nonextensive
approach. One can then study an ensemble in which the energy
($U$), temperature ($T$) and multiplicity ($N$), can all
fluctuate. These fluctuations are then connected by the following
relation:
\begin{eqnarray}
\Big| \omega^2_N - \frac{1}{\langle N\rangle}\Big| &=& \omega^2_U
+ \omega^2_T - 2\rho \omega_U \omega_T \label{eq:corq}\\
&& = \left( \omega_U - \omega_T\right)^2 + 2 \omega_U\omega_T(1 -
\rho) = |q - 1|,\nonumber
\end{eqnarray}
where $\rho = \rho (U,T) \in [-1,1]$ is the correlation
coefficient between $U$ and $T$. This generalizes Linhard's
thermodynamic uncertainty relation, Eq. (\ref{eq:JL}). The
correlation coefficient enters since when all variables, $U$, $N$
and $T$ fluctuate, the pairs of variables, $(U,N)$ and $(U,T)$,
cannot all be independent because
\begin{equation}
Var(U) = \langle T\rangle Cov(U,N) + \langle N\rangle Cov(U,T)
\label{eq:UTN}
\end{equation}
(cf., \cite{WWcov}). This means that, in general,
\begin{equation}
\omega_U = \rho(U,N) \omega_N + \rho(U,T) \omega_T .
\label{eq:general}
\end{equation}
where $\rho(X,Y)$ denotes the corresponding correlation
coefficients between variables $X$ and $Y$. It should be noticed
at this point that in the literature \cite{old} there is a similar
relation connecting the volume, $V$, pressure, $P$ and
temperature, $T$:
\begin{equation}
\omega^2_P = \omega^2_V + \omega^2_T, \label{eq:old}
\end{equation}
but we shall not discuss it here.

The observed systematics in energy dependence of the parameter
$q$, deduced from presently available data, is shown in Fig.
\ref{Fig_F_q}. From measurements of different observables one
observes that, for high enough energies, $ q > 1$ and that values
of $q$ found from different observables are different. The latter
is caused either by technical (methodical) problems or else by a
physical cause. The former arises when, for example, fluctuations
of temperature are deduced either from data averaged over other
fluctuations, or from more refined data also accounting for
fluctuations of other variables (as in \cite{WWW}, see Eq.
(\ref{eq:sumrule})). The latter case is connected with the fact
that the observed $q$'s were obtained in different parts of phase
space. In this case one gets an uncertainty relation
(\ref{eq:corq}) with the help of which one can connect
fluctuations observed in different parts of phase space. For
example, one can recalculate $q$ obtained from $P(N)$ (i.e.,
obtained from the whole phase space, see dashed line in Fig.
\ref{Fig_F_q}) and compare it with $q$  evaluated from $f(p_T)$
(i.e., obtained from only transverse part of phase space, see full
line in Fig. \ref{Fig_F_q})\footnote{See \cite{GTR} for details. A
comment is in order concerning results of Fig. \ref{Fig_F_q}
obtained from $f(y)$. Namely, it turns out that, in the fitting
procedure, parameters $T$ and $q$ are strongly correlated
\cite{qcompilation,WWW}. As a result $q$ values evaluated in
different analysis of rapidity distributions \cite{QLQT} differ
slightly from those presented here (they give $q$ values
comparable or somewhat higher that one obtained from multiplicity
distribution).}.

\begin{figure}[h]
\begin{center}
\resizebox{0.45\textwidth}{!}{
  \includegraphics{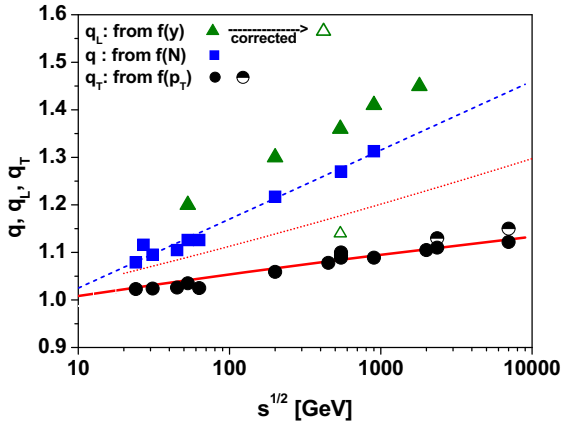}
  }
\caption{(Color online) Energy dependencies of the parameters $q$,
$q_L$ and $q_T$ as obtained from different observables. Triangles:
$q_L$ obtained from an analysis of rapidity distributions
\cite{RWW}; solid triangles show the uncorrected values, whereas
open triangle indicates the corrected value \cite{WWW}. Squares:
$q$ obtained from multiplicity distributions $P(N)$ (fitted by $q
= 1 + 1/k$ with $1/k = -0.104+0.029 \ln(s)$) \cite{PN}. Circles:
$q_T$ obtained from a different analysis of transverse momenta
distributions, $f\left(p_T \right)$. Data points in this case
come, respectively, from the \cite{PN} compilation of data (full
symbols) and from CMS data (half filled circles at high energies)
\cite{CMS}. The full and dotted lines come from Eq.
(\ref{eq:qqTqL}) and show, respectively, the energy dependence of
$q_T$ and energy dependence of $q_L$ (for $\rho = 0$, $\alpha =
2/3$ and $\kappa =1$).}. \label{Fig_F_q}
\end{center}
\end{figure}

\begin{figure}[h]
\begin{center}
\resizebox{0.45\textwidth}{!}{
  \includegraphics{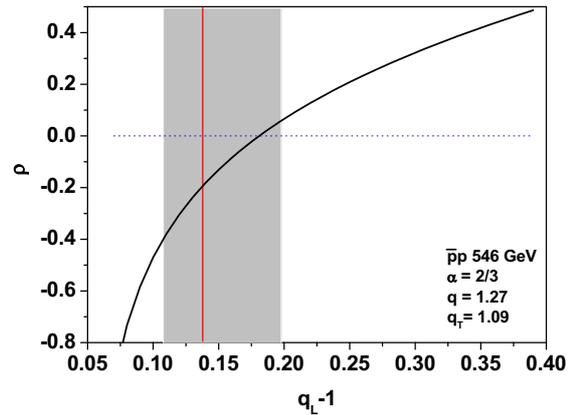}
  }
\caption{(Color online) Example of $\rho$ obtained from Eq.
(\protect\ref{eq:rho}). The shaded area shows the extent of
possible error, due to the uncertainty in fixing $q_L$.}
\label{F_rho}
\end{center}
\end{figure}
\begin{figure}[h]
\begin{center}
\resizebox{0.45\textwidth}{!}{
  \includegraphics{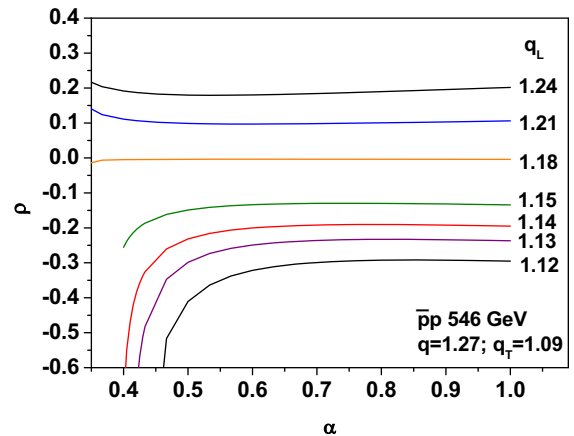}
  }
\caption{(Color online) Dependence of the correlation coefficient
$\rho$ on the parameter $\alpha$ for different values of $q_L$.}
\label{F_rho2}
\end{center}
\end{figure}

The correlation parameter $\rho$ appearing here bears important
information on the details of the production process. For example,
$\rho < 0$ means that a large energy $U$ (i.e., large inelasticity
of reaction, $K$) results in a large number of secondaries of
lower energies, whereas $\rho > 0$ means the opposite, one gets a
smaller number of larger energies. From Eq. (\ref{eq:corq}) one
finds that the coefficient $\rho$ is a function of all the
nonextensivity parameters involved. Denoting by $\alpha$ the part
of fluctuations of $T$ in the transverse direction, one finds
\begin{equation} q_T - 1 = \alpha \omega^2_T,\qquad q_L - 1 =
\omega^2_U + (1 - \alpha) \omega^2_T  \label{eq:qTL}
\end{equation}
and further
\begin{equation}
q - 1 = \left(q_L - 1\right) + \left(q_T - 1 \right) - 2 \rho
\omega_U \omega_T .\label{eq:qqTqL}
\end{equation}
It can be shown that
\begin{equation}
\kappa = \frac{\omega_U}{\omega_T} = \sqrt{\alpha \left( \frac{q_L
- 1}{q_T - 1} + 1 \right) - 1}. \label{eq:kovera}
\end{equation}
Finally, one obtains correlation coefficient $\rho$ expressed in
terms of different  fluctuations (in principle {\it measured})
(cf. \cite{PoS})\footnote{In \cite{WWcov} we used $\alpha =2/3$
and $\kappa = 1$; for $\rho = 0$. However, the actual values of
$\alpha$ and $\kappa$ parameters are irrelevant in this case.}:
\begin{equation}
\rho = \frac{ 1 - \frac{(q - 1) - \left( q_L - 1 \right)}{ q_T - 1
} }{ \frac{2}{\alpha} \sqrt{ \alpha\left( \frac{q_L - 1}{q_T - 1}
+ 1 \right) - 1 }};\qquad \alpha = \frac{q_T - 1}{\omega_T^2}.
\label{eq:rho}
\end{equation}

An example of the feasibility of deducing $\rho$ from data is
presented in Fig. \ref{F_rho} for data on $\bar{p} + p$ at $546$
GeV \cite{UA5}. In this case one takes from $P(N)$ $q = 1.27$,
from the distribution of $p_T$ one has $q_T = 1.09$, whereas from
the original $q_L = 1.36$ one obtains, after correction, $q_L =
1.14$ (cf. Fig. \ref{Fig_F_q}).

To summarize this part, note that, to get the correlation
coefficient $\rho$, one has to know {\it all the fluctuations},
i.e., both in the entire phase space, $q$, as separately in its
transverse, $q_T$, and longitudinal, $q_L$, parts. The best known
is $q$ (no corrections needed), for  $q_T$ the corrections are
small and can be neglected, finally, for $q_L$ the corrections are
large and must be accounted for (cf., Fig. \ref{F_rho}).

\subsection{Energy fluctuations - heat capacity}
\label{sec:Energy_fluct}

We now present energy fluctuations resulting from Tsallis
statistics and emerging from our analysis \cite{WWfluctU,WWcov}.
This subject already has its history (cf. \cite{LP}) and was also
recently under investigation (cf. \cite{Du}).

In Boltzman statistics \cite{WWfluctU} (with $kT = 1/\beta =
const$ and $N = const$) the energy $U = \sum^N_{i=1} E_i$ of $N$
particles is distributed according to
\begin{equation}
g_{T,N} = \frac{\beta}{\Gamma(N)}(\beta U)^{N - 1} \exp( - \beta
U) \label{eq:BoltU}
\end{equation}
for which
\begin{equation}
\frac{Var(U)}{\langle U\rangle^2} = \frac{k}{C^{(B)}_V} =
\frac{1}{N}\quad {\rm where}\quad C_V = \frac{\partial \langle
U\rangle}{\partial T}. \label{eq:CVB}
\end{equation}

In Tsallis statistics \cite{WWcov} one has, respectively,
\begin{eqnarray}
h_N(U) &=& \frac{\Gamma \left( N + \frac{2 - q}{q - 1}\right)}
              {\Gamma (N)\Gamma \left( \frac{2 - q}{q - 1}\right)}
              (q - 1)^N\beta ( \beta U )^{N - 1} \cdot\nonumber\\
      &\cdot& \left[ 1 - (1 - q)\beta U\right]^{\frac{1}{1
      - q} + 1 - N} \label{eq:TsallisU}
\end{eqnarray}
for which
\begin{eqnarray}
\frac{Var(U)}{\langle U\rangle^2} &=& \frac{1}{4 -
3q}\left( \frac{k}{C_V^{(T)}} + q - 1\right)  =\\
\label{eq:CVT} &=& \frac{1}{N} + \frac{q - 1}{4 - 3q}\left(1 +
\frac{1}{N}\right) \nonumber
\end{eqnarray}
where
\begin{equation}
C_V^{(T)} = \frac{\partial \langle U\rangle}{\partial T} =
Nk\frac{1}{3 - 2q} = C^{(B)}_V \frac{1}{3 - 2q}. \label{eq:CVT1}
\end{equation}
Notice that fluctuations of the energy $U$ are, in general, given
by the sum of two components: one obtained in the case of no
fluctuations and given by the heat capacity $C_V^{(B)}$ (which we
call the {\it kinetic component}) and one originating in
fluctuations, and given by the heat capacity $C^{(f)}$ (vanishing
when fluctuations vanish, we call it the {\it potential
component}):
\begin{equation}
\frac{Var(U)}{\langle U\rangle^2} = \frac{k}{C_V^{(B)}} +
\frac{k}{C^{(f)}} \label{eq:C_VBT}
\end{equation}
where (cf. \cite{LP})
\begin{equation}
C_V^{(B)} = kN\quad{\rm and}\quad C^{(f)} = k\frac{N}{N +
1}\frac{4 - 3q}{q - 1}. \label{eq:CBCTdef}
\end{equation}
From analysis of nuclear collisions we know \cite{WWprc} that $q$
depends on $N$:
\begin{equation}
q - 1 = \frac{\alpha}{N}, \label{eq:q(N)}
\end{equation}
where $\alpha$ is some constant of order unity depending on the
reaction considered. We can therefore write
\begin{eqnarray}
\frac{Var(U)}{\langle U\rangle^2} &=& \frac{1}{N} + \frac{q - 1}{4
- 3q}\left( 1 + \frac{1}{N}\right) = \nonumber\\
&=& \frac{1}{N} \left[ \frac{N(\alpha + 1) - 2\alpha}{N -
3\alpha}\right] \stackrel{N \rightarrow \infty}{\Longrightarrow}
0. \label{eq:CCCfin}
\end{eqnarray}
For small values of $q - 1$ (in practice already for $q - 1 <<
0.5$) one has
\begin{equation}
\frac{Var(U)}{\langle U\rangle^2} \ge (q - 1)\frac{1 +
\alpha}{\alpha} = \frac{1 + \alpha}{N}. \label{eq:finC}
\end{equation}

\section{Conditional probability - influence of conservation laws}
\label{section:Conditional}

Let $\{ E_{1,\dots,N}\}$ be a set of $N$ independent identically
distributed random variables described by some parameter $\lambda$
and let $g_N(E,\lambda)$ denote the gamma density function with
parameters $N$ and $\lambda$. For independent energies, $\{
E_{i=1,\dots,N}\}$, each distributed according to the simple
Boltzman distribution:
\begin{equation}
g_1\left(E_i\right) = \frac{1}{\lambda} \exp\left( -
\frac{E_i}{\lambda}\right), \label{eq:Boltzman}
\end{equation}
the sum
\begin{equation}
E = \sum^N_{i=1} E_i \label{eq:SumEi}
\end{equation}
is then distributed according to the following gamma distribution,
\begin{equation}
g_N(E) = \frac{1}{\lambda(N - 1)!}\left(
\frac{E}{\lambda}\right)^{N-1}\exp\left(
-\frac{E}{\lambda}\right). \label{eq:gammaE}
\end{equation}
If the available energy is limited, for example if $E =
\sum^N_{i=1} E_i = N \alpha = const$, then we have the following
conditional probability for the single particle distribution,
$f\left(E_i\right)$:
\begin{eqnarray}
f\left( E_i|E = N\alpha \right) &=& \frac{g_1\left(E_i\right)
g_{N-1}\left(N\alpha - E_i\right)}{g_N(N\alpha)} =\nonumber\\
&=& \frac{(N - 1)}{N}\frac{1}{\alpha}\left(1 - \frac{1}{N}
\frac{E_i}{\alpha} \right)^{N - 2}. \label{eq:condprob}
\end{eqnarray}
This is nothing else then the well known Tsallis distribution
\begin{equation}
f\left(E_i | E = const\right) = \frac{2 - q'}{\lambda} \left[ 1 -
(1 - q')\frac{E_i}{\lambda}\right]^{\frac{1}{1 - q'}}
\label{eq:condTsallis}
\end{equation}
with
\begin{equation}
q' = \frac{N - 3}{N - 2} < 1\quad {\rm and}\quad \lambda = \left(3
- 2q'\right)\alpha \label{eq:parcond}
\end{equation}
which is always less than unity. Here $\lambda = const$ and do not
fluctuate.

Now consider a situation in which the parameter $\lambda$ in the
joint probability distribution $$g\left( \{E_{1,\dots,N}\}\right)=
\prod^N_{i=1}g_i\left( E_i\right)$$ fluctuates according to a
Gamma distribution, Eq. (\ref{eq:gamma}). In this case we have the
single particle Tsallis distribution
\begin{equation}
h_i\left(E_i\right) = \frac{2 - q}{\lambda} \left[ 1 - (1 - q)
\frac{E_i}{\lambda}\right]^{\frac{1}{1 - q}} \label{eq:hTsallis}
\end{equation}
and the distribution of  $E=\sum^N_{i=1} E_i$  is given by (cf.
\cite{WWcov}):
\begin{eqnarray}
h_N(E) &=& \frac{(q - 1)^N \Gamma\left( N + \frac{2 - q}{q -
1}\right)}{\lambda \Gamma(N) \Gamma\left( \frac{2 - q}{q -
1}\right)}\cdot \nonumber\\
&&\cdot \left( \frac{E}{\lambda}\right)^{N - 1}\left[ 1 - (1 -
q)\frac{E}{\lambda} \right]^{1 - N + \frac{1}{1 - q}}.
\label{eq:hNE}
\end{eqnarray}
If  the energy is limited, i.e., if $E = \sum^N_{i=1} E_i = N
\alpha = const$,  we have the following conditional probability:
\begin{eqnarray}
f\left( E_i|E\right) &=& \frac{ h_i\left( E_i\right)h_{N-1}\left(E
- E_i\right)}{h_N(E)} = \nonumber\\
 &=& \frac{(N - 1)(2 - q)}{E[(3 - 2q) - N(1 -
 q)]}\frac{\lambda'}{\lambda}\left( \frac{E - E_i}{E}\right)^{N -
 1}\cdot \nonumber\\
 \cdot && \left[ 1 - (1 - q)\frac{E_i}{\lambda}\right]^{\frac{1}{1 -
 q}}\left[ 1 + (1 - q)\frac{E_i}{\lambda'}\right]^{2 - N + \frac{1}{1 -
 q}}\label{eq:fEiE}
\end{eqnarray}
where
\begin{equation}
\lambda' = \lambda - (1 - q)E. \label{eq:lambdap}
\end{equation}
For $q \rightarrow 1$ Eq. (\ref{eq:fEiE}) reduces to Eq.
(\ref{eq:condprob}). On the other hand, for large energy ($E
\rightarrow \infty$) and large multiplicity ($N \rightarrow
\infty$), the conditional probability distribution (\ref{eq:fEiE})
reduces to the single particle distribution given by
Eq.(\ref{eq:hTsallis}). Introducing the parameter $q'$ defined in
Eq.(\ref{eq:parcond}) the conditional probability (\ref{eq:fEiE})
can be rewritten as
\begin{eqnarray}
f\left( E_i|E\right)  &=& \frac{\left( 2 - q'\right)(2 - q)}{E[(3
- 2q)\left( 1 - q'\right) - \left( 3 - 2q'\right)(1 - q)]}\cdot
\frac{\lambda'}{\lambda} \cdot \nonumber\\
\cdot &&\left( \frac{E -
E_i}{E}\right)^{\frac{1}{ 1 - q'} }\cdot \nonumber\\
 \cdot && \left[ 1 - (1 - q)\frac{E_i}{\lambda}\right]^{\frac{1}{1 -
 q}}\left[ 1 + (1 - q)\frac{E_i}{\lambda'}\right]^{\frac{1}{1 - q} - \frac{1}{1 -
 q'}}.\label{eq:qprimq}
\end{eqnarray}
For $E_i << E$ it becomes
\begin{eqnarray}
f\left( E_i|E\right) &\simeq& \frac{\left(2 - q'\right)(2q - 1)(q
- 1)}{\lambda[(3 - 2q)\left(1 - q'\right) - \left(3 - 2q'\right)(1
- q)]}\cdot\nonumber\\
&\cdot& \left[1 - (1 - q)\frac{E_i}{\lambda}\right]^{\frac{1}{1 -
q}} \label{eq:eismallaE}
\end{eqnarray}
which, when additionally $N >> 1$ (or $q' \rightarrow 1$) reduces
to Eq. (\ref{eq:hTsallis}).

\begin{figure}[h]
\begin{center}
\resizebox{0.51\textwidth}{!}{
  \includegraphics{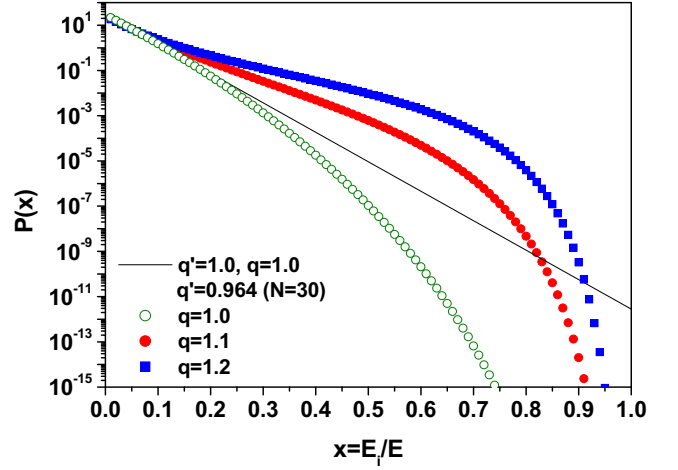}
  }
\caption{(Color online) Conditional probability distribution,
$P\left( x = E_i/E \right)$, for $q =1$ (
Eq.(\ref{eq:condTsallis})) and $q > 1$ (Eq.(\ref{eq:eismallaE}),
in both cases $N = 30$ ($q'= 0.964$), compared to exponential
distribution ($q = 1$, $q' = 1$).).
 } \label{Fig_conditional_1}
\end{center}
\end{figure}
\begin{figure}[h]
\begin{center}
\resizebox{0.51\textwidth}{!}{
  \includegraphics{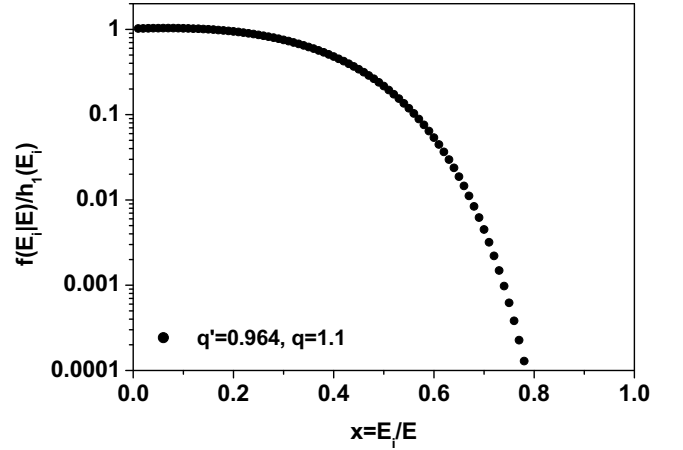}
  }
\caption{(Color online) Ratio of conditional distribution function
$f(E_i|E)$ and single particle distribution $h_1(E_i)$ as function
of $x = E_i/E$ for Tsallis statistics ($q = 1.1$ and $N = 30$).
 } \label{Fig_conditional_2}
\end{center}
\end{figure}

The results presented here are summarized in Figs.
\ref{Fig_conditional_1} and \ref{Fig_conditional_2} which shows
how large differences are (in $x = E_i/E$) between the {\it
conditional} Tsallis distribution $f(E_i|E)$ and the {\it usual}
$h_1(E_i)$\footnote{We would like to stress that
Eq.(\ref{eq:condprob}) has the form of a microcanonical
distribution in the one dimensional case, $D = 1$. In \cite{R2} it
was shown that smearing this distribution over a Gamma type
multiplicity distribution results in a microcanonical
generalization of the Tsallis distribution  which fits the
fragmentation functions measured in $e^+e^-$ experiments with
similar $q(s)$ evolution to that presented in Fig. \ref{Fig_F_q}.
It was demonstrated that this type of energy dependence seems to
be consistent with the DGLAP evolution equations \cite{R3}.}.

\section{Tsallis entropy and the Tsallis distribution function -
nonadditivity in nuclear collisions} \label{sec:Tsallis}

In all examples discussed so far we treated the Tsallis
distribution, Eq. (\ref{eq:Tsallis}), as a kind of superstatistics
\cite{SuperS} without really resorting to Tsallis entropy
\cite{Tsallis}. However, closer inspection of both approaches
reveals that the corresponding nonextensivity parameters (say $q$
and $q'$, respectively) are not identical. In fact one encounters
a sort of duality, like $q = 2 - q'$ discussed, for example, in
\cite{KGG,BJ,B}. We shall now address this problem in more detail
(cf., \cite{WWrev1} for details).

When starting from Tsallis entropy \cite{Tsallis},
\begin{equation}
S_q = \frac{1}{1 - q}\left[ \int dx f^q(x) - 1 \right],
\label{eq:T}
\end{equation}
one can obtain the probability density function $f(x)$ either by
optimizing it with constraints
\begin{equation}
\int dx f(x) = 1;\qquad  \int dx x f^q(x) = \langle x\rangle_q,
  \label{eq:xq}
\end{equation}
in which case \cite{RS}
\begin{eqnarray}
f(x) &=& (2 - q)\left[ 1 - (1 - q)x\right]^{\frac{1}{1-q}};
\label{eq:T1}\\
&&\qquad 0 \leq x < \infty;\quad 1\leq q\leq 3/2, \nonumber
\end{eqnarray}
or else by using as constraints
\begin{equation}
\int dx f(x) = 1;\qquad  \int dx x f(x) = \langle x\rangle,
\label{eq:x}
\end{equation}
in which case \cite{RS}
\begin{eqnarray}
f(x) &=& \frac{q}{\left[ 1 + (1 - q) x\right]^{\frac{1}{1-q}}};
\label{eq:T2}\\
&&\qquad 0 \leq x < \infty;\quad 1/2 < q\leq 1.
\end{eqnarray}
Out of these two possibilities, only (\ref{eq:T1}) is the same as
the distribution obtained in superstatistics and used above, cf.,
Eq. (\ref{eq:Tsallis}). On the other hand, the second
distribution, Eq. (\ref{eq:T2}), which seems to be more natural
from the point of view of a physical interpretation of the
constraint used, becomes the first one if expressed in terms of
$q'$ given by
\begin{equation}
q' = 2 - q . \label{eq:equivalence}
\end{equation}
Namely, in this case one has
\begin{equation}
f(x) = (2 - q')\left[ 1 - (1 - q')x\right]^{\frac{1}{1-q'}},
\label{eq:qprime}
\end{equation}
which, as show in Fig. \ref{Fig_F_q}, when compared to single
particle distributions, results in  $q' > 1$.

It turns out that there are data allowing the above duality (at
least in principle, considering the present status of their
quality). They are provided by nuclear collisions in which one
observes the apparent nonadditivities which, as will be shown,
allow us to compare and discuss both $q$ and $q'$\footnote{
Apparently similar duality occurs in nonextensive treatment of
fermions for which the particle-hole correspondence, $n_q(E, T,
\mu) = 1 - n_{2-q}(-E, T, -\mu)$ (where $\mu$ is the chemical
potential), must be preserved by the q-Fermi distributions
\cite{RW}. However, here we are facing different problem, namely
that parameter $q$ in entropy $S_q$ differs from parameter $q'$ in
probability distribution $f_{q'}$ with $q = 2 - q'$.}.

We start with the phenomenological approach used to describe
nuclear collisions which is based on the superposition model with
main ingredients being nucleons that have interacted at least once
\cite{WNM}. In this case, when sources are identical and
independent of each other, the total ($N$) and the mean ($\langle
N\rangle$) multiplicities are supposed to be given by,
\begin{equation}
N = \sum_{i=1}^{\nu}n_i, \qquad{\rm and}\qquad \langle N\rangle =
\langle \nu\rangle \langle n_i\rangle, \label{eq:Nu}
\end{equation}
where $\nu$ denotes the number of sources and $n_i$ the
multiplicity of secondaries from the $i^{th}$ source. Albeit at
present nuclear collisions are mostly described by different kinds
of statistical models \cite{MG_rev}, which automatically account
for possible collective effects, nevertheless a surprisingly large
amount of data can still be described by assuming the above
superposition of independent nucleon-nucleon collisions (possibly
slightly modified) as the main mechanism for the production of
secondaries. The question of the range of its validity is a
legitimate one \cite{FW}.

Using the notion of entropy, and considering $\nu$ independent
systems for which the corresponding individual probabilities are
combined as
\begin{equation}
p^{(\nu)}_q\left(x_1,\dots,x_{\nu}\right) =
\prod_{k=1}^{\nu}p^{(1)}_q\left( x_k\right), \label{eq:prodnu}
\end{equation}
and assuming that all $p^{(1)}_q\left(x_k\right)$ are the same for
all $k$ (i.e., their corresponding entropies $S_q^{(1)}$ are
equal), one finds
\begin{eqnarray}
S_q^{(\nu)} &=& \sum^{\nu}_{k=1}\frac{\nu !}{(\nu - k)!k!}(1 -
q)^{k - 1}\left[ S^{(1)}_q\right]^k =\nonumber\\
&=& \frac{\left[ 1 + (1 - q) S^{(1)}_q \right]^{\nu} - 1}{1 - q}.
\label{eq:Snu}
\end{eqnarray}
Notice that
\begin{equation}
\ln \left[ 1 + (1 - q) S^{(\nu)}_q\right] = \nu \ln \left[ 1 + (1
- q) S^{(1)}_q\right] \label{eq:lnS}
\end{equation}
and that
\begin{equation}
S^{(\nu)}_q \stackrel{ q \rightarrow 1}{\longrightarrow} \nu \cdot
S^{(1)}_1. \label{eq:lnSln} \
\end{equation}
For $q < 1$ one has
\begin{equation}
\frac{S^{(\nu)}_q}{\nu} \stackrel{\nu \rightarrow
\infty}{\longrightarrow} \infty, \label{eq:Snu}
\end{equation}
i.e., entropy $S^{(\nu)}_q$ is nonextensive. For $ q > 1$ one has
\begin{equation}
S^{(\nu)}_q \ge 0\quad{\rm  only~ for}\quad q < 1 +
\frac{1}{S^{(1)}_q}
\end{equation}
and
\begin{equation}
\frac{S^{(\nu)}_q}{\nu} \stackrel{\nu \rightarrow
\infty}{\longrightarrow} 0,
\end{equation}
 i.e., entropy is extensive,
 \begin{equation}
 0 \leq \frac{ S^{(\nu)}_q}{\nu} \le S^{(1)}_q.
\end{equation}
In the following we put $\nu = N_W/2 = N_P$ ($N_W$ is the number
of wounded nucleons and $N_P$ is the number of participants from a
projectile). Assuming naively that the total entropy is
proportional to the mean number of produced particles,
\begin{equation}
S = \alpha \langle N\rangle , \label{eq:SN}
\end{equation}
one obtains the following relation between mean multiplicities in
$AA$ and $NN$ collisions,
\begin{equation}
\alpha \langle N\rangle_{AA} = \frac{\left[1 + (1 - q)
\alpha\langle N\rangle_{pp}\right]^{N_P} - 1}{1 - q}.
\label{eq:alphaN}
\end{equation}
At this point we stress the following observation, so far not
discussed in detail. Namely, because (as shown in \cite{WWprc}),
$\langle N\rangle_{AA}$ increases nonlinearly with $N_P$ and
$\langle N\rangle_{AA} > N_P \cdot \langle N\rangle_{pp}$, the
nonextensivity parameter obtained here from considering the
corresponding entropies must be smaller than unity, $q < 1$. On
the other hand, all estimations of the nonextensivity parameter
(let us denote it by $q'$) discussed before lead to $q' > 1$. This
is the {\it $q$ duality in nonextensive statistics} mentioned
above, on which we shall concentrate in more detail.

To start with, the relation (\ref{eq:alphaN}) is not exactly
correct for $S_q$. In what follows we denote entropy on the level
of particle production by $s$ (and the corresponding
nonextensivity parameter by $\tilde{q}$), whereas the
corresponding entropies and nonextensivity parameter on the level
of $NN$ collisions by $S$ and $q$. From Eq. (\ref{eq:Snu}) we have
that for $N$ particles
\begin{equation}
s^{(N)}_{\tilde{q}} = \frac{\left[ 1 + \left(1  -
\tilde{q}\right)s^{(1)}_{\tilde{q}} \right]^N - 1}{1 - \tilde{q}}
\stackrel{\tilde{q} \rightarrow 1}{\longrightarrow} N\cdot
s^{(1)}_{\tilde{q}} = \alpha N, \label{eq:tildeq}
\end{equation}
where $s^{(1)}_{\tilde{q}} = \alpha$ is the entropy of a single
particle. In a $A+A$ collision with $\nu$ nucleons participating,
Eq. (\ref{eq:Snu}) results in
\begin{equation}
S^{(\nu)}_{q} = \frac{\left[ 1 + \left(1  - q\right)S^{(1)}_{q}
\right]^{\nu} - 1}{1 - q}, \label{eq:justq}
\end{equation}
where $S^{(1)}_q$ is the entropy of a single nucleon.

Denoting multiplicity in a single $N+N$ collision by $n$, the
respective entropy is
\begin{equation}
S^{(1)}_q = S^{(1)}_{\tilde{q}} = \frac{\left[ 1 + \left(1 -
\tilde{q}\right)s^{(1)}_{\tilde{q}} \right]^n - 1}{1 - \tilde{q}},
\label{eq:single}
\end{equation}
whereas the entropy in a $A+A$ collision for $N$ produced
particles is
\begin{equation}
S^{(N)}_{\tilde{q}} = \frac{\left[ 1 + \left(1 -
\tilde{q}\right)s^{(1)}_{\tilde{q}} \right]^N - 1}{1 - \tilde{q}}.
\label{eq:AA}
\end{equation}
This means that
\begin{equation}
S^{(N)}_{\tilde{q}} = S^{(\nu)}_q. \label{eq:SS}
\end{equation}
Notice that parameters $q$ and $\tilde{q}$ are usually not
identical. Moreover, from the relation
\begin{equation}
 q - 1 = \frac{1}{aN_P}\left( 1 - \frac{N_P}{A}\right),\qquad a =
 \frac{C_V}{N_P}  \label{eq:qN_P}
\end{equation}
one finds that for $NN$ collisions (where $N_P=A$) $\tilde{q} =
1$. On the other hand, for $\tilde{q} = q$ Eq. (\ref{eq:SS})
corresponds to the situation encountered in superpositions, as in
this case one has
\begin{equation}
\left[ 1 + (1 - q)s^{(1)}_q\right]^N = \left[1 + (1 -
q)s^{(1)}_q\right]^{n\nu}  \label{eq:Nnun}
\end{equation}
and so
\begin{equation}
N= n\nu .\label{eq:n_nu}
\end{equation}

Consider now the general case and denote
\begin{equation}
c_1 = 1 + \left(1 -
\tilde{q}\right)s^{(1)}_{\tilde{q}};\qquad\qquad c_2 =
\frac{1-q}{1 - \tilde{q}}. \label{eq:c1c2}
\end{equation}
These quantities are not independent because:
\begin{equation}
c_2 c_1^N + 1 - c_2 = \left( c_2 c_1^n + 1 -
c_2\right)^{\nu}.\label{eq:ccc}
\end{equation}
From relation (\ref{eq:ccc})
\begin{equation}
\frac{N}{\nu\cdot n} = \frac{1}{\nu n\cdot \ln c_1} \ln\left[
\frac{\left(c_2c_1^n + 1 - c_2\right)^{\nu} - \left( 1 -
c_2\right)}{c_2}\right], \label{eq:FB}
\end{equation}
which for $N = \langle N_{AA}\rangle $, $n = \langle
N_{pp}\rangle$ and $\nu = N_P$ is presented in Fig. \ref{Fig_BC}
for different reactions. As seen there one can describe
experimental data by using $c_2 = 1.7$ and with $c_1$ depending on
energy $\sqrt{s}$ according to $c_1(s) = 1.0006 - 0.036
s^{-1.035}$. Notice that for energies $\sqrt{s} > 7$ GeV one has
$c_1 > 1$. This means that $\tilde{q} < 1$ and (because $c_2 > 0$)
also $ q < 1$.
\begin{figure}[h]
\begin{center}
\resizebox{0.51\textwidth}{!}{
  \includegraphics{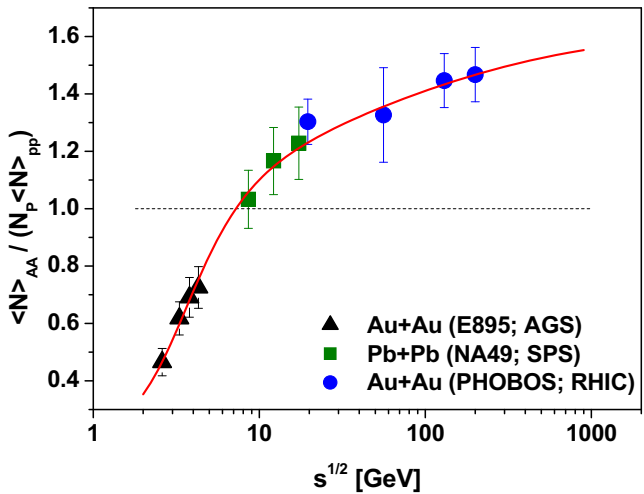}
  }
\caption{(Color online) Energy dependence of the charged
multiplicity for nucleus-nucleus collisions divided by the
superposition of multiplicities from proton-proton collisions (cf.
Eq. (\ref{eq:FB})). Experimental data on multiplicity are taken
from the compilation \cite{PHOBOS}.} \label{Fig_BC}
\end{center}
\end{figure}

To summarize this section, we have shown that, non additivity in
the superposition model described using the notion of entropy
clearly requires $q < 1$, cf. Figs. \ref{Fig_BC}. This means that
$q'$ is not the same as $q$. The conclusion one can derive from
these considerations is that the second way of deriving $f(x)$,
which uses a linear condition, cf. Eq. (\ref{eq:T2}), is the
correct one, and that $q'$ in the distribution is not the same as
$q$ in the entropy. The problem is that, whereas from
distributions one can easily deduce a numerical value of $q'$,
this is not the case when one uses entropy (at least not when
deduced from presently available data). There are too many
variables to play with (cf., considerations using the
superposition model as above). For example, in the definition of
$c_1$ in Eq. (\ref{eq:c1c2}), one has $s^{(1)}_{\tilde q}$, which
is not known {\it a priori}. The only thing one can deduce in this
case is that $q < 1$. We cannot therefore check numerically that
relation (\ref{eq:equivalence}) really holds. But, if one agrees
that the Tsallis distribution comes from Tsallis entropy, we have
only two options: either $q' = q$ or $q' -1 = 1 -q$. Our
conclusion presented here, that $q' > 1$ and $q < 1$, therefore
supports the second option, i.e., Eq. (\ref{eq:equivalence}).

However, this final observation calls for comment. Namely, the
probability density function (PDF) is usually evaluated by the
Maximum Entropy Method (MEM) for Tsallis entropy with some
constraints \cite{OM} \footnote{Notice that Tsallis entropy is a
monotonic function of the Renyi entropy, $S_q =
\ln_q\left[\exp\left(R_q\right)\right]$, and both lead to the same
equilibrium statistics of particles (with coinciding maxima in
equilibrium for similar constraints on the expectation value).}.
Therefore the situation is not unique since there are four
possible well documented MEMs \cite{E6} using two kinds of
definition for an expectation value of the physical quantities:
the normal average (\ref{eq:x}) and the $q$-average (\ref{eq:xq})
(with normal, as here, or the so-called escort PDFs
\cite{escort}). Although various arguments justifying it  have
been given \cite{E11} it was also been pointed out that, for a
small change of the PDF, thermodynamic averages obtained by the
$q$-averages are unstable, whereas those obtained by the normal
average are stable \cite{E14}. On yet another hand, it is claimed
that for the escort PDF, the Tsallis entropy and thermodynamical
averages are robust \cite{E17} . All this means that the stability
(robustness) of thermodynamical averages as well as the Tsallis
entropy is still a controversial issue \cite{E18}.

\section{Examples of nonfluctuating (nonthermal) mechanisms leading to Tsallis
distribution} \label{section:nonthermal}

It should be realized that the so far discussed origins of the
Tsallis distribution, based either on superstatistics or on
Tsallis entropy, are by no means the only possibilities. Therefore
we end with short discussions of two examples of obtaining Eq.
(\ref{eq:Tsallis}) in a completely nonthermal way, these are the
application of {\it order statistics} and the use of {\it
stochastic networks}.

\subsection{Order statistics} \label{section:Order}

Order statistics is based on the observation \cite{WWorder} that
the selection of the minimal value of the ordered variables leads
in a natural way to its distribution being given Eq.
(\ref{eq:Tsallis}) (with $q$ both greater and smaller than unity,
depending on circumstances), i.e., in fact by the Tsallis
distribution , the same as that resulting from Tsallis
nonextensive statistics. Distribution of the minimal values of
some specific choices of the variable $E$ is known in the
literature as order statistics \cite{Order})\footnote{Actually,
one can easily invent a nonthermal scenario leading to a
thermal-like form of the observed spectra, see, for example,
recent work \cite{CKS}. In such an approach the resultant
distribution emerges not because of the equilibration of energies
due to some collisions (i.e., because of the kinematic
thermalization), but rather because of the process of erasing of
memory of the initial state and is the result of the approaching
to a state of maximal entropy (called in \cite{CKS} stochastic
thermalization).}.

\begin{figure}[h]
\begin{center}
\resizebox{0.5\textwidth}{!}{
  \includegraphics{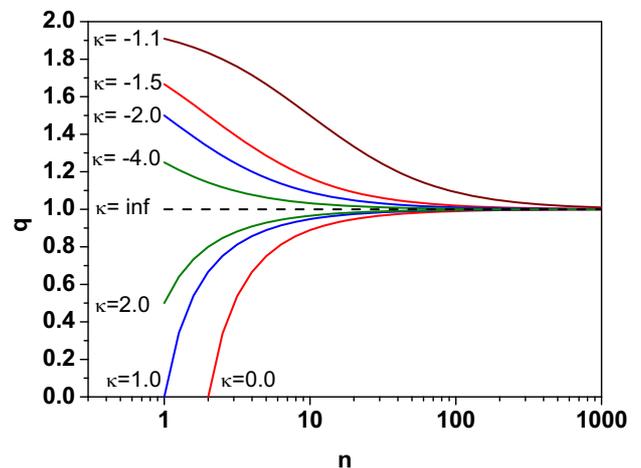}
  }
\caption{(Color online) $q$ as function of $n$ given by Eq.
(\ref{eq:qorder}) for  different values of $\kappa$.}
\label{Fig_order}
\end{center}
\end{figure}

We now present a generalized version of what was proposed in
\cite{WWorder}. We start with a set of $n$ virtual particles (so
called {\it ghost-particles}) with energies $\varepsilon_i$ taken
from some distribution $f\left(\varepsilon \right)$. Ordering the
values of $\varepsilon_i$ (i.e., introducing in this set {\it rank
statistics}), $\varepsilon_1 < \varepsilon_2 < \dots <
\varepsilon_n$, we choose a {\it real particle} with minimal
energy $E = \varepsilon_1 = \min\left( \{ \varepsilon_i\}\right)$.
It is straightforward to find a function $g(E)$ describing the
energy distribution of real particles. The probability density to
find a particle with energy $E$ among $n$ elements is $n f(E)$.
The probability to find particles with energy greater than $E$ is
$1 - F(E)$, where $F(E) = \int^E_0d\varepsilon f(\varepsilon)$ is
the distribuant of $f$ If a particle of energy $E$ is already that
of the minimal energy it means that the remaining $n-1$ particles
have to poses  higher energies. The probability of such an event
is equal to $[1 - F(E)]^{n-1}$. This means that the distribution
of the minimal value in sample of $n$ elements is\footnote{More
formally, the cumulative distribution function is $G(E) = 1 - [1 -
F(E)]^n$ and the density distribution is $g(E) = dG(E)/dE = n[1 -
F(E)]^{n - 1}dF(E)/dE = nf(E)[1 - F(E)]^{n - 1}$. }
\begin{equation}
g(E) = nf(E)[1- F(E)]^{n-1} . \label{eq:g(E)}
\end{equation}
Because $f(E) = d F(E)$, the distribution $g(E)$ is properly
normalized if $f(E)$ is normalized. For
\begin{equation}
f(\varepsilon) = - \alpha (\kappa + 1)(1 + \alpha
\varepsilon)^{\kappa} \label{eq:fvarepsilon}
\end{equation}
where $\kappa \neq - 1$ (because of the normalization requirement)
and $\alpha = - sign (\kappa + 1) \beta$ ($\beta = 1/T > 0$) one
gets $g(E)$ in the form of Tsallis distribution, Eq.
(\ref{eq:Tsallis}), with
\begin{equation}
q = \frac{n(\kappa + 1) - 2}{n(\kappa + 1) -1}, \label{eq:qorder}
\end{equation}
$q > 1$ for $\kappa < - 1$ and $q < 1$ for $\kappa > - 1$.   Fig.
\ref{Fig_order} shows $q(n)$ dependence for different values of
the parameter $\kappa$ (special cases of $\kappa = - 2$ and
$\kappa = 0$ were discussed in \cite{WWorder}).

\subsection{Stochastic networks}
\label{sec:networks}

Stochastic network structures occur in almost all branches of
modern science (including sociology and economy). They have
therefore been the subject of intensive research, also by means of
Tsallis statistics (cf. \cite{WWnet,Tnet} for details and full
list of references; in \cite{WWnet1} this approach has been
applied to multiparticle production processes\footnote{In
\cite{WWnet1} the "power laws", assumed {\it ad hoc} in
\cite{MGMG} (as a kind of opposition to Tsallis statistics), was
explained using a stochastic networks approach presented here.
Actually, this "power laws" idea is continued recently in
\cite{recentPL} as an apparent new observation. It must be
mentioned therefore that this idea is actually quite old; such a
type of parametrization of $p_T$ distributions has been proposed
(and was shown to be phenomenological successful) already in
\cite{CM}.}). There are two basic types of stochastic networks:
\begin{itemize}
\item Networks with a constant number of nods, $M$, for which
probability that given node has $k$ connections with other nodes
($k$ links) is poissonian \cite{ER},
\begin{equation}
P(k) = \frac{\kappa_0^k}{k!}\cdot e^{-\kappa_0};\qquad \kappa_0 =
\langle k\rangle . \label{eq:ER}
\end{equation}
\item Networks in which the number of nodes is not stationary and
the distribution of links $P(k)$ is given by dynamics of the
growth of network \cite{BAJ}. It varies between being {\it
exponential},
\begin{equation}
P(k) = \frac{1}{\kappa}\cdot \exp\left( - \frac{k}{\kappa}\right),
\label{eq:equal}
\end{equation}
and {\it power-like},
\begin{equation}
P(k) = \frac{2\kappa^2t}{\kappa_0+t}\cdot k^{-3},
\label{eq:prefer}
\end{equation}
behavior. In the former case each new node connects with the
already existing ones with equal probability, $\Pi(k_i)
=1/(\kappa_0+t-1)$, independent of $k_i$. In the latter case one
has preferential attachment (the so called "rich-get-richer"
mechanism, here $\kappa < \kappa_0$ is the number of new nodes
added in each time step) with, in this case,
$\Pi(k_i)=k_i/(2\kappa t)$ choice.
\end{itemize}

Let us remind ourselves that, whereas
\begin{equation}
\frac{df(x)}{dx} = - \frac{1}{\lambda}f(x)\quad
\Longrightarrow\quad f(x) = \frac{1}{\lambda} \exp\left( -
\frac{x}{\lambda}\right), \label{eq:Tnet1}
\end{equation}
for the $x$-dependent scale parameter
\begin{equation}
\lambda \rightarrow \lambda(x) = \lambda_0 - (q - 1) x
\label{eq:Tnet2}
\end{equation}
the exponential solution takes a power-like form,
\begin{equation}
f(x) = \frac{2 - q}{\lambda_0}\left[ 1 - (1 -
q)\frac{x}{\lambda_0}\right]^{\frac{1}{1 - q}}. \label{eq:Tnet3}
\end{equation}
For preferential attachment used in \cite{WWnet}, dividing the
"master equation"
\begin{equation}
\frac{\partial P(k)}{\partial t} = - c P(k) \label{eq:Tnet4}
\end{equation}
by the assumed "growth of the network"
\begin{equation}
\frac{\partial k}{\partial t} = a + bk, \label{eq:Tnet5}
\end{equation}
one obtains the following evolution equation for the network
considered:
\begin{equation}
\frac{\partial P(k)}{\partial k} = - c P(k) \frac{\partial
t}{\partial k} = - \frac{c}{a + bk}P(k). \label{eq:Tnet6}
\end{equation}
For $c = 1$, $a = \kappa_0$ and $b = q - 1$ one has $\kappa(k) =
\kappa_0 + (q - 1)k$ and a solution of Eq. (\ref{eq:Tnet6}) in the
form of the Tsallis distribution, Eq. (\ref{eq:Tnet3}):
\begin{equation}
P(k) = \frac{2 - q}{\kappa_0}\left[ 1 - (1 -
q)\frac{k}{\kappa_0}\right]^{\frac{1}{1 - q}}.\label{eq:Tnet7}
\end{equation}
For $q\rightarrow 1$ Eq. (\ref{eq:Tnet7}) recovers Eq.
(\ref{eq:equal}) whereas for $k >> \kappa_0/(q-1)$ it leads to
"scale-free" power distribution
\begin{equation}
P_q(k) \propto k^{- \gamma},\quad {\rm with}\quad \gamma =
\frac{1}{q - 1}. \label{eq:Tnet8}
\end{equation}
The frequently observed value $\gamma = 3$ therefore corresponds
to $q = 4/3$. At this value of $q$ the variance of distribution
$P(k)$ diverges,
\begin{equation}
Var(k) = \frac{\kappa_0^2(2 - q)}{(3 - 2q)^2(4 - 3q)}\quad
\stackrel{q \rightarrow 4/3}{\Longrightarrow} \infty.
\label{eq:Tnet9}
\end{equation}

We close this section by noticing that formally we can interpret
Eq. (\ref{eq:Tnet6}) as the stationary solution, of the following
Fokker-Planck equation,
\begin{equation}
\frac{d\left(K_2P(k)\right)}{dk} = K_1 P(k),\label{eq:FP1}
\end{equation}
where $ K_1 = q - 2$ and $K_2 = \kappa_0 + (q - 1)k$, This
corresponds (cf. network growth given by Eq. (\ref{eq:Tnet5})) to
the Langevin equation with multiplicative noise($\eta$) in the
form \cite{BJ}:
\begin{equation}
\frac{\partial k}{\partial t} + \eta k = \xi, \label{eq:L}
\end{equation}
where $\xi$ is the traditional noise term. In this case both
noises have nonzero mean values: $\langle \xi(t)\rangle =
\kappa_o$ and $\langle \eta(t)\rangle = 1 - q$, and correlations:
$Cov\left(\xi(t),\xi\left( t'\right)\right) = 2\kappa_0
\delta\left(t - t'\right)$,
$Cov\left(\eta(t),\eta\left(t'\right)\right)=0$ and $Cov\left(
\eta(t),\xi\left(t'\right)\right) = (1-q)\delta\left(t-t'\right)$.

\section{Summary}
\label{sec:Summary}

The possibility of occurrence of intrinsic, nonstatistical
temperature fluctuations has far-reaching consequences which we
have attempted to present in this review (covering results
obtained since \cite{WWrev} or not covered there but worth
mentioning). Our work in this field started with a realization
that in a nonhomogeneous heat bath one can expect some heat
diffusion process to operate. This then results in specific
fluctuations of the temperature $T$, eventually resulting in a
Tsallis distribution Eq. (\ref{eq:Tsallis}) \cite{WWq}.
Notwithstanding vivid discussions concerning the legitimacy of
such a possibility \cite{FLUCT}, this idea has been further
elaborated and generalized in \cite{WWrev,BJ,B}.

The results presented here can be summarized as follows:
\begin{itemize}
\item Fluctuations of $T$ (of any kind) result in Tsallis
distributions (\ref{eq:Tsallis}) with $q > 1$.

\item Observables from different parts of phase space are
characterized by different values of $q - 1$. We understand why
this is so and are able to connect $q$ as obtained from different
observables.

\item Constraints imposed by the conservation laws result in a
distortion of the Tsallis distribution. In the limiting case (when
unconditional distributions are of BG type) conditional
distributions become of the Tsallis type with $q < 1$.

\item Tsallis distributions with $q > 1$ correspond to Tsallis
entropy with $q' < 1$.

\item The so called "power law", propositions which occur in the
literature \cite{CM,MGMG,recentPL}, are nothing else but Tsallis
distributions {\it in disguise}.

\end{itemize}

\begin{acknowledgement}
Acknowledgment: Partial support (GW) of the Ministry of Science
and Higher Education under contract DPN/N97/CERN/2009 is
gratefully acknowledged. We would like to warmly thank Dr Eryk
Infeld for reading this manuscript.
\end{acknowledgement}

\end{document}